\documentstyle{aipproc}
\def\etal{{\it et al.}}

\begin{document}
\title{A Statistical Treatment of the Gamma-Ray Burst ``No Host
Object'' Issue} 

\author{David L. Band$^{1}$ and Dieter H. Hartmann$^{2}$}
\address{$^{1}$CASS, UC San Diego, La Jolla, CA 92093 \\
$^{2}$Dept. of Physics \& Astronomy, Clemson University, Clemson, SC 29634}

%\lefthead{LEFT head}
%\righthead{RIGHT head}
\maketitle

\begin{abstract}
Various burst origin scenarios require a host object in the burst
error box or near a well-located position.  For example, a host galaxy
should be present in the standard cosmological models.  We present a
methodology which evaluates whether the observed detections and
nondetections of potential host objects in burst error boxes are
consistent with the presence of the host, or whether all the
detections can be attributed to background objects (e.g., unrelated
background galaxies).  The host object's flux distribution must be
modeled.  Preliminary results are presented for the ``minimal''
cosmological model. 
\end{abstract}
\section*{Introduction}
In many gamma-ray burst scenarios a host object should be detected
when an error box is observed to sufficiently faint fluxes. However,
once an error box has been observed, how do we know whether the host
object has been detected?  Most cosmological models predict that
bursts occur in or near galaxies.  Since the study of X-ray and
optical transients indicate that some and probably all bursts are
cosmological, here we will focus on galaxies as the host objects,
although the concepts and methodology can be applied to other host
object types. 

The study of burst error boxes consists of three interrelated aspects.
 First is the observations of the error boxes, which we assume result
in a list of galaxies which are brighter than a limiting flux.  These
observations can be in any wavelength band in which imaging is
possible, although usually optical or infrared images are used. Second
is the model for the host object, which guides both the observations
and the analysis.  For example, the assumption that bursts are
cosmological leads the observer to ignore the stars in the error box,
although the observer (hopefully) notices any unusual objects in the
field.  Third is the analysis of the observations in terms of the
model.  Beyond \hbox{deciding} whether the observations support the
model, the analysis methodology also guides the observer as to which
error boxes should be searched and to what detection limit. Here we
present a new methodology for analyzing burst error box observations,
and present preliminary results.  A more complete presentation is in
press\cite{band98}. 

We emphasize that the analysis of burst error boxes must be made in
the context of a model of the expected host.  It is nonsensical to ask
merely whether galaxies are present in an error box because if one
searches deep enough one will find a multitude of faint galaxies. Here
we test the ``minimal'' cosmological model used in most studies of
burst ensembles, particularly those analyzing burst error boxes. 
Bursts are assumed to be standard candles in this model:  a basic
burst property such as total emitted energy or peak photon luminosity
is constant for all bursts, and does not evolve with redshift.
Therefore, there is a one-to-one mapping between a burst's redshift
and the observed intensity corresponding to the standard candle (e.g.,
energy fluence for a constant total emitted energy); the
redshift-intensity relationship is derived from the intensity
distribution under the assumption that the comoving density of burst
sources does not evolve.  Of course, in this model bursts occur in
galaxies.  Since a neutron star-neutron star merger is a possible
origin of a burst's energy, and the number of compact binary systems
is presumably proportional to a galaxy's mass, the burst rate per
galaxy is assumed to be proportional to the galaxy's luminosity (for a
constant mass-to-light ratio)\cite{fenimore93}. Undoubtedly bursts are
characterized by a luminosity function, and cosmological density and
luminosity evolution is likely, but this ``minimal'' model has been a
reasonable working assumption in the absence of additional data. 

B.~Schaefer\cite{schaefer92} first reported that the galaxies in 8 
burst error boxes were fainter than expected.  Specifically, Schaefer 
calculated a large burst energy (up to $2\times10^{53}$ ergs) if the 
brightest galaxy in an error box was as bright as M31.  Fenimore 
\etal\cite{fenimore93} introduced the statistic
\begin{equation}
S=\int_0^{f_{\rm det}} df \, \psi(f)
\end{equation}
where the brightest galaxy in an error box has a flux of $f_{\rm det}$
and $\psi(f)$ is the flux distribution of the expected host galaxies;
$S$ is the fraction of the distribution which is fainter than $f_{\rm
det}$. If $f_{\rm det}$ is indeed the flux of the host galaxy, and
$\psi(f)$ is the correct distribution, then $S$ should be distributed
uniformly between 0 and 1, with an average of 1/2$\pm (12N)^{-1/2}$
for $N$ error boxes (this test is similar to the V/V$_{\rm max}$
test).  Based on the minimal cosmological model, Fenimore \etal\ found
$\langle S \rangle=0.44\pm 0.10$ for Schaefer's data. While this
$\langle S \rangle$ is consistent with the minimal model, the value of
$S$ for a given error box is only an upper limit since the brightest
galaxy may be a background galaxy instead of the host galaxy.
Similarly, although S.~Larson\cite{larson96,larson97a,larson97b} reported
an overabundance of bright galaxies in his K-band observations of nine 
IPN$^3$ boxes, he recognized that many of these galaxies are
unrelated background galaxies. 

We therefore derived an analysis methodology which includes the
unrelated background galaxies in evaluating whether a host galaxy is
present.  An additional guiding principle was the use of all available
information.  Thus the method uses all the detected galaxies in the
observations.  We describe the error box by a probability density
$\rho(\Omega)$, where $\Omega$ represents the spatial coordinates.
Typically it is assumed that $\rho(\Omega)= 1/\Omega_0$ within the
99\% contour (a region of size $\Omega_0$), and $\rho(\Omega)=0$
outside, but more sophisticated treatments are possible.  The
detection threshold may vary across the error box, e.g., as a result
of mosaicing the box with multiple observations of differing quality. 
Currently we do not include the clustering of background galaxies,
which should be a small effect. 
\section*{Methodology}
Our method is a Bayesian comparison of two hypotheses: $H_0$---a host
galaxy is present in addition to unrelated background galaxies; and
$H_1$---only background galaxies are present.  Assume the observations
of an error box reveal $n_d$ galaxies with fluxes $f_i$ above a
detection limit $f_{\rm lim}(\Omega)$; these results we represent by
the statement $D$.  We set up an odds ratio 
\begin{equation}
O(H_0,H_1) = {{p(H_0 \,|\, D)}\over{p(H_1 \,|\, D)}}
   ={{p(H_0)}\over{p(H_1)}} {{p(D \,|\, H_0)}\over{p(D \,|\, H_1)}}
\end{equation}
where: $p(H_x \,|\, D)$ is the probability that hypothesis $H_x$ is
true given the observations $D$; $p(H_x)$ is the ``prior,'' our
assessment of the validity of $H_x$ before obtaining the new data $D$;
and $p(D \,|\, H_x)$ is the likelihood of $H_x$, the probability of
obtaining $D$ if the hypothesis $H_x$ is correct.  For simplicity we
set the two priors equal to each other, $p(H_0)=p(H_1)$.  Therefore,
the odds ratio is the ratio of the likelihoods, $O(H_0,H_1) = p(D
\,|\, H_0)/p(D \,|\, H_1)$. 

The likelihoods are calculated by breaking into little bins the
three-dimensional space formed by the two spatial dimensions and the
flux, and calculating the probability that a host or background galaxy
is present or absent in each bin.  If the galaxy redshifts are also
available, then the redshift can be added as a fourth dimension. 
Poisson statistics characterize the  probability that a background
galaxy is found in a given bin.  The likelihood for $H_0$, $p(D \,|\,
H_0)$, is the sum of every possibility for the presence of a host
galaxy: either the host is fainter than the limit $f_{\rm lim}$ or it
is one of the detected galaxies.  Consequently\cite{band98} 
\begin{equation}
{{p(D \, | \, H_0)} \over {p(D \, | \, H_1) }} =
   \int d\Omega \int_0^{f_{\rm lim}(\Omega)} df \,\Psi(f)\rho(\Omega) 
%   \nonumber \\
   + \sum_{i=1}^{n_d} {{\Psi(f_i,z_i)\rho(\Omega_i)}
   \over {\phi(f_i,z_i) }}
\end{equation}
where $\Psi(f,z)$ is model-dependent host galaxy distribution,
$\rho(\Omega)$ is the burst location's probability density across the
error box, $\phi(f,z)$ is distribution of background galaxies, and
$n_d$ is the number of detected galaxies.  If the redshifts of the 
detected galaxies are unknown, then the $z$-dependence of $\Psi$ and 
$\phi$ should be dropped.  In this equation, the first
term on the right is the probability that the host galaxy can be
hidden below the detection limit, while the sum compares for each
detected galaxy the probability that it is the host galaxy to the
probability that it is an unrelated background galaxy.  Clustering of 
the background galaxies can be included by multiplying $\phi(f_i,z_i)$ 
by a function of the distance to the other detected galaxies.  This 
additional factor will usually be of order unity, and will not affect 
our results qualitatively.

For a database with a number of error boxes the likelihood ratio for
the ensemble is the product of the ratios for each box.  If the
resulting odds ratio is much larger than one, then the presence of
host galaxies has been demonstrated.  If the ratio is much less than
one, then the host galaxy model is incorrect.  Finally, if the ratio
is of order unity, then the data are insufficient to distinguish
between the hypotheses. 

By evaluating the likelihood ratio for the expected host and
background galaxies, we can determine the method's sensitivity for a
given error box.  For reasonable assumptions about the distributions,
we find that an observation can distinguish between hypotheses if the
error box is small enough so that the expected host galaxy is much
brighter than average background galaxy. Only then is it clear that a
galaxy is the host and not a background galaxy.  As currently
formulated, this methodology tests a given hypothesis. However, it can
easily be modified to fit model parameters. 

This methodology was developed for finite size error boxes, such as
has been available from the various IPNs.  However, the methodology
can be readily adapted for other circumstances.  Afterglows localize
the burst with very small uncertainties (e.g., a fraction of an
arcsecond).  However, unless the model being tested places the burst
in a galactic nucleus, a galaxy within a certain region around the
burst would be acceptable as the host; this region can be treated as
the error box.  Similarly, the burst source might have been ejected
from the host galaxy.  The distance the source might have traveled
before bursting can be used to define the error box around an
afterglow; finite size error boxes should be expanded by this
distance. 
\section*{Applications}
As examples, we apply this methodology to published datasets to test
the ``minimal'' cosmological model described above.  In the future we
plan to test variants of the cosmological model using a more extensive
dataset. 

Larson and McLean\cite{larson97b} presented K-band observations of 9
IPN$^3$ error boxes with an average size of 8~arcmin$^2$.  They listed
only the flux of the brightest galaxy in each box, and therefore we
use this galaxy as the single galaxy detection and its flux as the
detection limit. For all 9 error boxes we find 
\begin{equation}
\prod_{j=1}^{9} O_j = 0.25
\end{equation}
which indicates that based on this data we cannot determine whether or
not a host galaxy is present.  The reason the analysis of these data
is inconclusive is that the fluxes of the average expected host
galaxy, the detection limit, and the average brightest background
galaxy are all comparable; therefore even if the host galaxy is
present, the odds ratio will be of order unity. 

Schaefer \etal\cite{schaefer97} observed 4 small (1/4-2~arcmin$^2$)
burst error boxes with the {\it Hubble Space Telescope}. The detection
of objects exhibiting bizarre behavior (e.g., proper motion) was the
primary purpose of these observations, but our methodology can be
applied to the data, nonetheless.  Galaxies were detected in 2 of
these error boxes.  The odds ratio for the four boxes together is 
\begin{equation}
\prod_{j=1}^{4} O_j = 2\times10^{-6}\quad .
\end{equation}
This is a clear statement that host galaxies expected by the minimal 
model are not present.
\section*{Acknowledgements}
We thank B.~Schaefer, C.~Luginbuhl and F.~Vrba for stimulating
discussions.  This research is supported by the {\it CGRO} guest
investigator program (DLB and DHH) and NASA contract NAS8-36081 (DLB).

\end{document}